\documentclass[fleqn,10pt]{article}
\usepackage[utf8]{inputenc}
\usepackage[T1]{fontenc}

\usepackage{a4wide}
\usepackage{graphicx}
\usepackage{amsmath, amssymb}
\usepackage{psfrag}

\newcommand{\mR}{\mathbb{R}}

\newcommand{\X}{\mathcal{X}}

\usepackage{multicol}
\usepackage[superscript,nomove]{cite}
\usepackage{newtxtext}
\usepackage{newtxmath}

\title{\bfseries Memcapacitors and Meminductors are\\ Overunity Systems!}

\author{Dimitri Jeltsema\thanks{School of Engineering and Automotive, HAN University of Applied Science, P.O.~Box 2217, 6802 CE Arnhem, The Netherlands. Email: d.jeltsema@han.nl} \and Arjan van der Schaft\footnote{Bernoulli Institute for Mathematics, Computer Science and AI, University of Groningen, P.O.~Box 407, 9700 AK Groningen, The Netherlands. Email: a.j.van.der.schaft@rug.nl}}

\date{}

\begin{document}

\maketitle

\begin{abstract}
It is rigorously proved that ideal memcapacitors and meminductors are \emph{not} passive or lossless devices, nor are they satisfying the weaker notion of cyclo-passivity, which arises when dropping the requirement of non-negativity of the storage function. Equivalently, this implies that there exist excitation profiles that allow to extract more energy from the device than it is supplied with; so that their energy conversion efficiency exceeds 100\%. This means that ideal memcapacitors and meminductors constitute so-called \emph{overunity} systems. An illustrative mechanical analogue is provided that explicitly confirms this property. Hence, the question arises if ideal memcapacitors and meminductors will just remain some mathematical toys or artefacts.   
\end{abstract}

\section{Introduction}\label{sec:intro}

Batteryless circuit elements that are able to store information would represent a serious paradigm change in electronics, allowing for low-power computation and storage. One such candidate circuit element is the ideal \emph{memristor},\cite{Chua1971} which gained a worldwide attention of both researchers and the mainstream media over the past decade since its supposed physical realization was announced.\cite{Strukov2008} The notion of memelements can also be extended to ideal capacitive and inductive devices, coined as \emph{memcapacitors} and \emph{meminductors}.\cite{DiVentra2009} Generalizations to include any class of two-terminal devices whose resistance, capacitance, and inductance depends on the internal state of the system are known as memristive, memcapacitive, and meminductive systems (memsystems for short), respectively. Both ideal memelements and generalized memsystems have specific properties that typically appear most strikingly as a pinched hysteretic loop in the two constitutive variables that define them: current versus voltage for memristive systems, voltage versus charge for memcapacitive systems, and current versus flux-linkage for the meminductive systems. 

Many real-world systems are argued to belong to the class of memsystems, especially those of nanoscale dimensions.\cite{Chua2003,DiVentra2009,Radwan2015}. However, to the best of our knowledge, and apart from the fact that already the real-world existence of an ideal memristor is critically addressed,\cite{Vongehr2015,Abraham2018} physical devices that either constitute an ideal memcapacitor or an ideal meminductor are yet to be found. 

It is well-known that real-world devices generally cannot store more energy than they are supplied with. This means that such devices are generally \emph{passive} systems. At most they can be locally active, which means that at some interval within a given time-frame energy might be created but will be lost again at another interval within the same time-frame. Such devices are referred to as \emph{cyclo-passive} systems. 

The main contribution of this paper is to prove that memcapacitors and meminductors are neither passive nor cyclo-passive. Non-passive devices that are able to produce `free' energy are known as so-called \emph{overunity} systems or, in a more classical parlance, \emph{perpetual motion} systems. Hence, the possibility of inventing or discovering real-world ideal energy-storing memelements is highly unlikely. 

The remainder of the paper is organized as follows. First, the mathematical essentials of ideal memcapacitors and meminductors are collected in Section \ref{sec:E-memelements}. Section \ref{sec:diss_theory} briefly recalls the theory of dissipativity and cyclo-dissipativity, which is instrumental in Section \ref{sec:memcap_memind_passivity}
 to analyze the (cyclo-)passivity properties of the ideal memcapacitor and meminductor. The outcome of the analysis is exemplified in Section \ref{sec:mech_analog} by using a mechanical analogue that is recently proposed in the literature. Some concluding remarks are provided in Section \ref{sec:conclusion}. 

\section{Energy-Storing Memelements}\label{sec:E-memelements}

Let $V$ and $I$ denote the port voltage and current, respectively, and their respective time-integrals, $\varphi$ and $q$, the flux-linkage and charge, i.e., 
\begin{equation*}
\varphi(t) :=\varphi(t_0) + \int\limits_{t_0}^{t} V(\tau)d\tau \ \text{and} \ q(t) := q(t_0) + \int\limits_{t_0}^{t} I(\tau)d\tau,
\end{equation*}  
or, equivalently, $\dot{\varphi} = V$ and $\dot{q} = I$. 

\subsection{Memcapacitor} 

An ideal memcapacitor\cite{DiVentra2009} is defined by a constitutive relationship between flux-linkage $\varphi$ and time-integrated charge $\rho$, i.e.,
\begin{equation}\label{eq:constit_memcap}
\rho = \hat{\rho}(\varphi)  \ \text{or} \  \varphi = \hat{\varphi}(\rho).
\end{equation}
Differentiating the latter with respect to time yields the following dynamical systems.
\begin{multicols}{2}
\noindent Voltage-controlled memcapacitor:
\begin{equation}\label{eq:V-memcap}
\begin{aligned}
\dot{\varphi} &= \frac{q}{C(\varphi)},\\
\dot{q} &= I \quad \text{(input)},\\[0.5em]
\text{(output)} \quad V &= \frac{q}{C(\varphi)}, 
\end{aligned}
\end{equation}
where $C(\varphi):=\dfrac{d\hat{\rho}}{d\varphi}(\varphi)$ denotes the capacitance. 

\noindent Charge-controlled memcapacitor:
\begin{equation}\label{eq:Q-memcap}
\begin{aligned}
\dot{\rho} &= q,\\[0.5em]
\dot{q} &= I \quad \text{(input)},\\[0.5em]
\text{(output)} \quad V &= K(\rho)q, 
\end{aligned}
\end{equation}
where $K(\rho):=\dfrac{d\hat\varphi}{d\rho}(\rho)$ denotes the electrical elastance (inverse capacitance). 

\end{multicols}

\subsection{Meminductor}

An ideal meminductor\cite{DiVentra2009} is defined by a constitutive relationship between charge $q$ and time-integrated flux-linkage $\sigma$, i.e.,
\begin{equation}\label{eq:constit_memind}
\sigma = \hat{\sigma}(q)  \ \text{or} \  q = \hat{q}(\sigma).
\end{equation}
Differentiating the latter with respect to time yields the following dynamical systems.
\begin{multicols}{2}
\noindent Current-controlled meminductor:
\begin{equation}\label{eq:I-memind}
\begin{aligned}
\dot{q} &= \frac{\varphi}{L(q)},\\
\dot{\varphi} &= V \quad \text{(input)},\\[0.5em]
\text{(output)} \quad I &= \frac{\varphi}{L(q)}, 
\end{aligned}
\end{equation}
where $L(q):=\dfrac{d\hat{\sigma}}{dq}(q)$ denotes the inductance. 

\noindent Flux-controlled meminductor:
\begin{equation}\label{eq:P-memind}
\begin{aligned}
\dot{\sigma} &= \varphi,\\[0.5em]
\dot{\varphi} &= V \quad \text{(input)},\\[0.5em]
\text{(output)} \quad I &= \varGamma(\sigma)\varphi, 
\end{aligned}
\end{equation}
where $\varGamma(\sigma):=\dfrac{d\hat{q}}{d\sigma}(\sigma)$ denotes the inverse inductance. 
\end{multicols}

\section{Dissipativity and Cyclo-Dissipativity Theory}\label{sec:diss_theory}

Let us first recall the basic definitions of dissipativity theory as originating from the seminal work of Willems\cite{willems1972}, see also Hill \& Moylan\cite{hillmoylan1980} and Van der Schaft\cite{passivitybook}, and then turn attention to the less well-known, and weaker, notion of cyclo-dissipativity. 

Consider a system with state vector $x$ and a vector of external (e.g., input and output) variables $w$. Consider a scalar-valued {\it supply rate} $s(w)$. A function $S(x)$ is said to be a {\it storage function} (with respect to the supply rate $s$) if along all trajectories of the system and for all $t_1 \leq t_2$ and $x(t_1)$ it satisfies the {\it dissipation inequality}
\begin{equation}\label{diss}
S\big(x(t_2)\big) - S\big(x(t_1)\big) \leq \int\limits_{t_1}^{t_2} s(w(t)) dt.
\end{equation}
Interpreting $s\big(w(t)\big)$ as `power' supplied to the system at time $t$, and $S\big(x(t)\big)$ as stored `energy' while at state $x(t)$, this means that increase of the stored energy can only occur due to externally supplied power. 

Following Willems,\cite{willems1972} the system is called {\it dissipative} (with respect to the supply rate $s$) if there exists a {\it non-negative} storage function $S$. [Since addition of an arbitrary constant to a storage function again leads to a storage function, the requirement of non-negativity of $S$ can be relaxed to $S$ being {\it bounded from below}.] Furthermore, it is called {\it lossless} (with respect to $s$) if there exists a non-negative storage function satisfying the dissipation inequality \eqref{diss} with {\it equality}.

An external characterization of dissipativity, in terms of the behavior of the $w$ trajectories, is the following.\cite{willems1972} Define for any $x$ the expression (possibly infinite)
\begin{equation}
S_a(x):= \sup_{w, T\geq 0} - \int\limits_0^T s\big(w(t)\big) dt,
\end{equation}
where the supremum is taken over all external trajectories $w(\cdot)$ of the system corresponding to initial condition $x(0)=x$, and all $T\geq 0$. Obviously, $S_a(x)\geq 0$. Then the system is dissipative with respect to the supply rate $s$ if and only if $S_a(x)$ is {\it finite} for every $x$. 

Interpreting as above $s(w)$ as `power' supplied to the system, $S_a(x)$ is the maximal `energy' that can be extracted from the system at initial condition $x$, and  the system is dissipative if and only if this maximal `energy' is finite for any $x$. Furthermore, if  $S_a(x)$ is finite for every $x$ then $S_a$ is itself a non-negative storage function, and is in fact the {\it minimal} non-negative storage function (generally among many others). 

Dropping the requirement of non-negativity of the storage function $S$ leads to the notion of {\it cyclo-dissipativity}, respectively {\it cyclo-losslessness}.\cite{willems1973, hillmoylan1975, passivitybook, vdscyclo}. First note that if a general, possibly indefinite, storage function $S$ satisfies \eqref{diss}, then for all {\it cyclic} trajectories, i.e., $x(t_1)=x(t_2)$, we have that
\begin{equation}\label{cyclic}
\int\limits_{t_1}^{t_2}  s\big(w(t)\big) dt \geq 0,
\end{equation}
which holds with equality in case \eqref{diss} holds with equality. This leads to the following external characterization of {\it cyclo}-dissipativity and {\it cyclo}-losslessness.

\bigskip

\noindent {\bfseries Definition.}
{\it 
A system is {\it cyclo-dissipative} if (\ref{cyclic}) holds
for all $t_2\geq t_1$ and all external trajectories $w$ such that $x(t_2)=x(t_1)$. In case \eqref{cyclic} holds with equality, we speak about cyclo-losslessness. Furthermore, the system is called cyclo-dissipative {\it with respect to} $x^*$ if \eqref{cyclic} holds for all $t_2\geq t_1$ and all external trajectories $w$ such that $x(t_2)=x(t_1)=x^*$, and cyclo-lossless with respect to $x^*$ if this holds with equality.
}

\bigskip

Interpreting again $s(w)$ as power provided to the system, cyclo-dissipativity thus means that for any cyclic trajectory the net amount of energy supplied to the system is non-negative, and zero in case of cyclo-losslessness.

The following theorem\cite{vdscyclo} extends the results in Hill \& Moylan\cite{hillmoylan1975} and shows the equivalence between the external characterization of cyclo-dissipativity and cyclo-losslessness and the existence of storage functions. 

\bigskip

\noindent {\bfseries Theorem.}
{\it
If there exists a function $S$ satisfying the dissipation inequality \eqref{diss}, then the system is cyclo-dissipative, and it is cyclo-lossless if $S$ satisfies \eqref{diss} with equality.
Conversely, assume the system is reachable from some 
ground-state $x^*$ and controllable to this same state $x^*$. [It is immediate that this property is independent of the choice of $x^*$.] Define the (possibly extended) functions $S_{ac}: \X \to \mR \cup \infty$ and $S_{rc} : \X \to - \infty \cup \mR$ as
\begin{equation}
\begin{aligned}
S_{ac}(x) &= \!\!\!\! \mathop{\sup_{w,T\geq 0}}_{x(0)=x,x(T)=x^*} - \int\limits_0^Ts(w(t)) dt, \\[2mm]
S_{rc}(x) &= \!\!\!\!\!\! \mathop{\inf_{w,T\geq 0}}_{x(-T)=x^*,x(0)=x}   \ \ \int\limits_{-T}^0s(w(t)) dt,
\end{aligned}
\end{equation}
where the supremum and infimum are taken over all external trajectories $w(\cdot)$ and $T\geq 0$.
Then the system is cyclo-dissipative with respect to $x^*$ if and only if
\begin{equation}\label{acrc}
S_{ac}(x) \leq S_{rc}(x), \ \text{for all} \ x \in \X.
\end{equation}
Furthermore, if the system is cyclo-dissipative with respect to $x^*$ then both $S_{ac}$ and $S_{rc}$ are storage functions, and thus the system is cyclo-dissipative. Finally, $S_{ac}(x^*) = S_{rc}(x^*)=0$, and any other storage function $S$ satisfies
\begin{equation*}
S_{ac}(x) \leq S(x) - S(x^*) \leq S_{rc}(x),
\end{equation*}
while if the system is cyclo-lossless $S_{ac}(x)= S(x) - S(x^*)= S_{rc}(x)$, implying uniqueness (up to a constant) of the storage function.
}

\bigskip

Note that, unlike the dissipativity case, it may be possible for a cyclo-dissipative system to extract an {\it infinite} amount of energy from the system (since the storage function may not be bounded from below).

Finally, consider an input-state-output system 
\begin{equation*}
\begin{aligned}
\dot{x} & = f(x,u),\\
y & = h(x,u),
\end{aligned}
\end{equation*}
with state variables $x \in \X$, input(s) $u$, and output(s) $y$, with $w=(u,y)$. Then for any {\it differentiable} storage function $S$ the dissipation inequality \eqref{diss} is equivalent to its differential version\cite{willems1972, hillmoylan1980, passivitybook}
\begin{equation*}
\frac{\partial S}{\partial x}(x)f(x,u) \leq s\big(u,h(x,u)\big), \ \text{for all } \ x,u.
\end{equation*}

In case of the {\it passivity} supply rate $s(w)=s(u,y)=y^Tu$, where $w=(u,y)$ and $u$ and $y$ are equally dimensioned vectors, the terminology `dissipativity' in all of the above is replaced by the classical terminology of {\it passivity}. In this case $u$ and $y$ typically are vectors of power-conjugate variables, like forces and velocities, and voltages and currents. 
For an input-affine system of form
\begin{equation*}
\begin{aligned}
\dot{x} & = f(x) + g(x)u,\\
y & = h(x),
\end{aligned}
\end{equation*}
with state variables $x \in \X$, input(s) $u$, and output(s) $y$, cyclo-passivity amounts to the existence of a storage function (nonnegative in the case of passivity) $S:\X \to \mathbb{R}$ such that\cite{hillmoylan1975,hillmoylan1980,passivitybook}
\begin{equation}\label{eq:passivity_test}
h(x) = g^T(x)\frac{\partial S}{\partial x}(x)  \ \text{and} \  \left[\frac{\partial S}{\partial x}(x)\right]^T \! f(x) \leq 0. 
\end{equation}
This characterization of (cyclo-)passivity will be instrumental in analysing the (cyclo-)passivity properties of the ideal memcapacitor and meminductor in the next section. 

\section{Ideal Energy-Storing Memelements are Not (Cyclo-)Passive}\label{sec:memcap_memind_passivity}

Consider the charge-controlled memcapacitor (\ref{eq:Q-memcap}). Suppose that $S(\rho,q)$ is a storage function with respect to the passivity supply rate $s(I,V)=IV$. Then, according to (\ref{eq:passivity_test}), the system (\ref{eq:Q-memcap}) is (cyclo-)passive if and only if the following two properties are satisfied:
\begin{subequations}
\begin{align}
&\frac{\partial S}{\partial q}(\rho,q) = K(\rho) q,\label{eq:equality_Q_memcap}\\
&\frac{\partial S}{\partial \rho}(\rho,q)q \leq 0.\label{eq:inequality_Q_memcap}
\end{align}
\end{subequations}
From the equality (\ref{eq:equality_Q_memcap}) it follows that
$S(\rho,q) = \frac{1}{2}K(\rho)q^2 + G(\rho)$, for some function $G(\rho)$. Substituting the latter into the inequality (\ref{eq:inequality_Q_memcap}) yields that
\begin{equation*}
\frac{1}{2}K'(\rho)q^3 + G'(\rho)q \leq 0,
\end{equation*}
for all $\rho$ and $q$, and $(\cdot)'$ denotes the ordinary derivative with respect to the function's argument. However, this readily implies that $G'(\rho)=0$ (and thus that $G$ is a constant function) and $K'(\rho) = \hat{\varphi}''(\rho)=0$, which implies that $\varphi$ can at most be an affine function of $\rho$, i.e., $\varphi = K_0 +  K_1\rho$.
Hence the output equals $V=K_1 q$, which just constitutes an ordinary \emph{linear} capacitor. In conclusion, a charge-controlled memcapacitor (\ref{eq:Q-memcap}) is \emph{not} passive nor cyclo-passive!

Performing the same analysis as above for the voltage-controlled memcapacitor (\ref{eq:V-memcap}), we see that $S(\varphi,q)$ is a storage function if and only if 
\begin{equation*}
\frac{\partial S}{\partial q}(\varphi,q) = \frac{q}{C(\varphi)} \ \Leftrightarrow \ S(\varphi,q) = \frac{q^2}{2C(\varphi)}+G(\varphi),
\end{equation*}
and
\begin{equation*}
\frac{\partial S}{\partial \varphi}(\varphi,q)\dot{\varphi} \leq 0 \ \Leftrightarrow \ \left[-\frac{q^2}{2C^2(\varphi)}C'(\varphi) + G'(\varphi) \right]\frac{q}{C(\varphi)} = - \frac{C'(\varphi)}{2C^3(\varphi)}q^3 + \frac{G'(\varphi)}{C(\varphi)}q\leq 0,
\end{equation*}
which, using $q = C(\varphi)V$, implies that
\begin{equation*}
\frac{1}{2}C'(\varphi)V^3 + G'(\varphi)V \leq 0,
\end{equation*}
for all $\varphi$ and $V$. Like before, the latter inequality can only be satisfied if and only if $C'(\varphi) = G'(\varphi) = 0$. This means that the only admissible constitutive relationship must be linear, i.e., $\rho = C_0 + C_1 \varphi$, which again implies an ordinary linear capacitor $q=C_1 V$, and $G$ is again any constant function. 

Thus, ideal memcapacitors, either charge- or voltage-controlled, are \emph{not} passive nor cyclo-passive! Moreover, the same conclusion follows \emph{mutatis-mutandis} for both the ideal meminductor systems (\ref{eq:P-memind}) and (\ref{eq:I-memind}). By the theory of Section \ref{sec:diss_theory} this implies that for \emph{any} nonlinear constitutive relationship in (\ref{eq:constit_memcap}) or (\ref{eq:constit_memind}) there should \emph{always} exists an input function that let us extract more energy from the device than it is supplied with. This will be exemplified in the next section. 

\section{Overunity Devices?}\label{sec:mech_analog}

As mentioned in Section \ref{sec:intro}, although a variety of so-called generalized candidate memcapacitive and meminductive \emph{systems} have been proposed in the literature,\cite{DiVentra2009,Radwan2015} an ideal real-world electrical memcapacitor or meminductor satisfying (\ref{eq:V-memcap})-(\ref{eq:Q-memcap}) or (\ref{eq:I-memind})-(\ref{eq:P-memind}) has not been found yet. The reason is rather obvious, as illustrated by the following mechanical analogy.  

\subsection{Mechanical Analogue: The Mem-Inerter}

Adopting the Firestone (mobility) analogy \cite{Jeltsema2009} for which voltage is considered analogous to velocity and current is analogous to force, suggests that the voltage-controlled memcapacitor (\ref{eq:V-memcap}) can be considered as the electrical analogue of a recently proposed and experimentally realized mechanical displacement-dependent \emph{mem-inerter}.\cite{Zang2018,Zang2020} Such system exhibits a displacement-dependent inertance (mechanical `capacitance') of the form
\begin{equation}\label{eq:mass}
B(z) = b_0\left(\frac{w}{2}-z\right),
\end{equation}
where $z \in [-w/2,w/2]$ denotes the relative displacement (mechanical `flux-linkage') of the piston, and $b_0$ and $w$ are the base inertance  and the working width of the piston, respectively. Hence, we obtain the following dynamical system (compare with (\ref{eq:V-memcap}))
\begin{equation}\label{eq:inerter}
\begin{aligned}
\dot{z} & = \frac{p}{B(z)},\\
\dot{p} &= F \quad \text{(input)},\\
\text{(output)} \quad v &= \frac{p}{B(z)},
\end{aligned}
\end{equation}
where $p$ represent the linear momentum (mechanical `charge'), $F$ the applied force (mechanical `current'), and $v=\dot{z}$ the associated velocity (mechanical `voltage'). 

\bigskip

\noindent {\bfseries Remark.}
{\it
Alternatively, adopting the classical Maxwell-Kelvin analogy \cite{Jeltsema2009} for which voltage is considered analogous to force and current is analogous to velocity, the mem-inerter constitutes an equally valid mechanical analogue of a current-controlled meminductor (\ref{eq:I-memind}).
}

\subsection{Example of `Free' Energy Harvesting}

Consider the mem-inerter (\ref{eq:inerter}), which, as shown in Section \ref{sec:memcap_memind_passivity}, is not passive nor cyclo-passive. This implies that there exists an input signal for which we can extract more energy from the system than it is supplied with. Indeed, adopting the same parameters as in Zang et.~al.\cite{Zang2018} yields $b_0=9381.7$ [kg] and $w=0.1$ [m], and selecting for instance the following (zero-mean) periodic force profile
\begin{equation*}
F(t) = 2.5\sin(\omega t) - 0.25\cos(\omega t) -5\sin(2\omega t) - 1.25\cos(2\omega t) + 2.75\cos(3\omega t) - 1.25\cos(4\omega t) \ \text{[N]},
\end{equation*}
with $\omega=0.5$ [rad/s] and fundamental period time $T=2\pi/\omega$ [s]. 

\begin{figure*}[t]
\begin{center}
\includegraphics[width=0.5\textwidth]{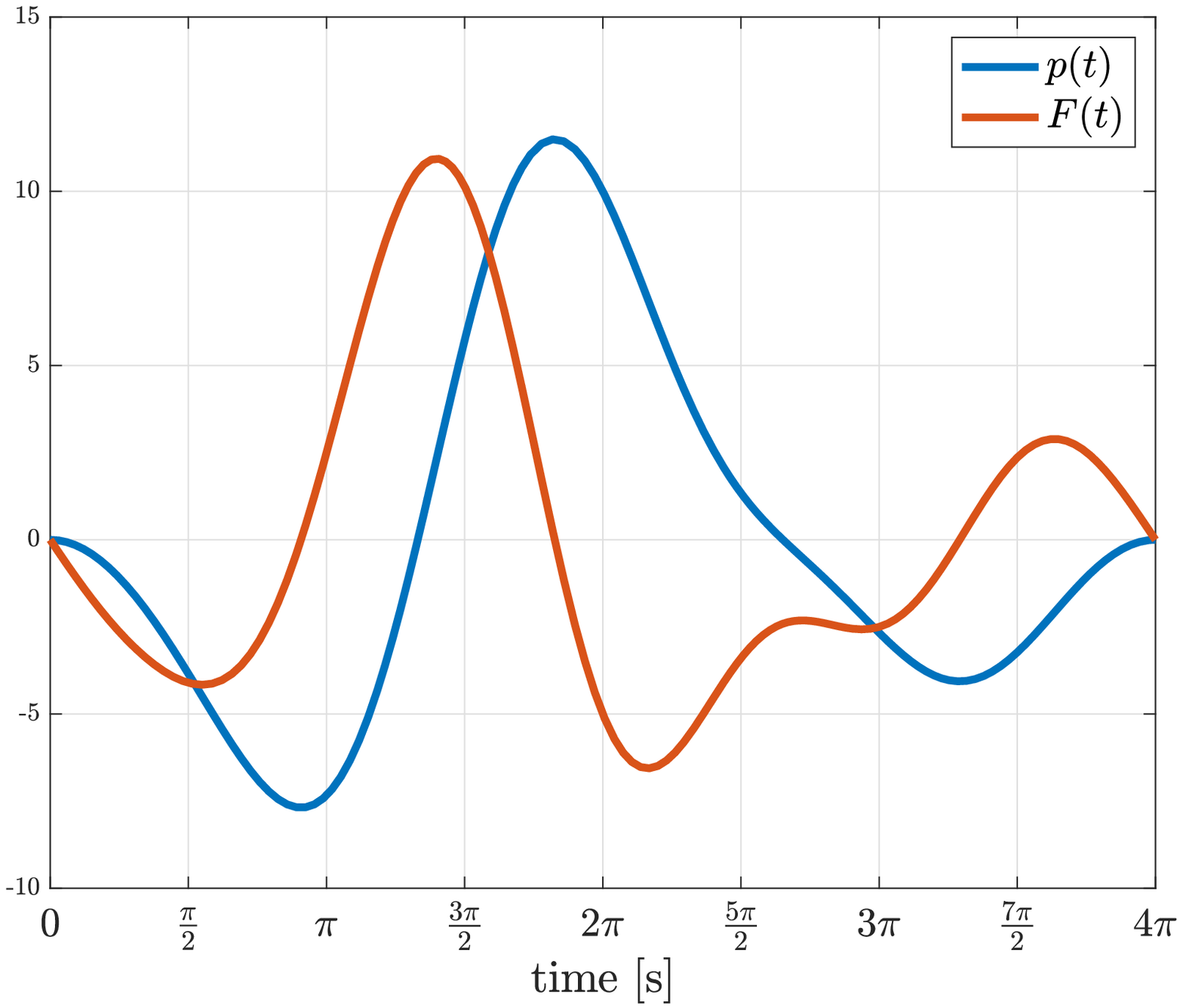}\hspace*{-5mm}
\includegraphics[width=0.5\textwidth]{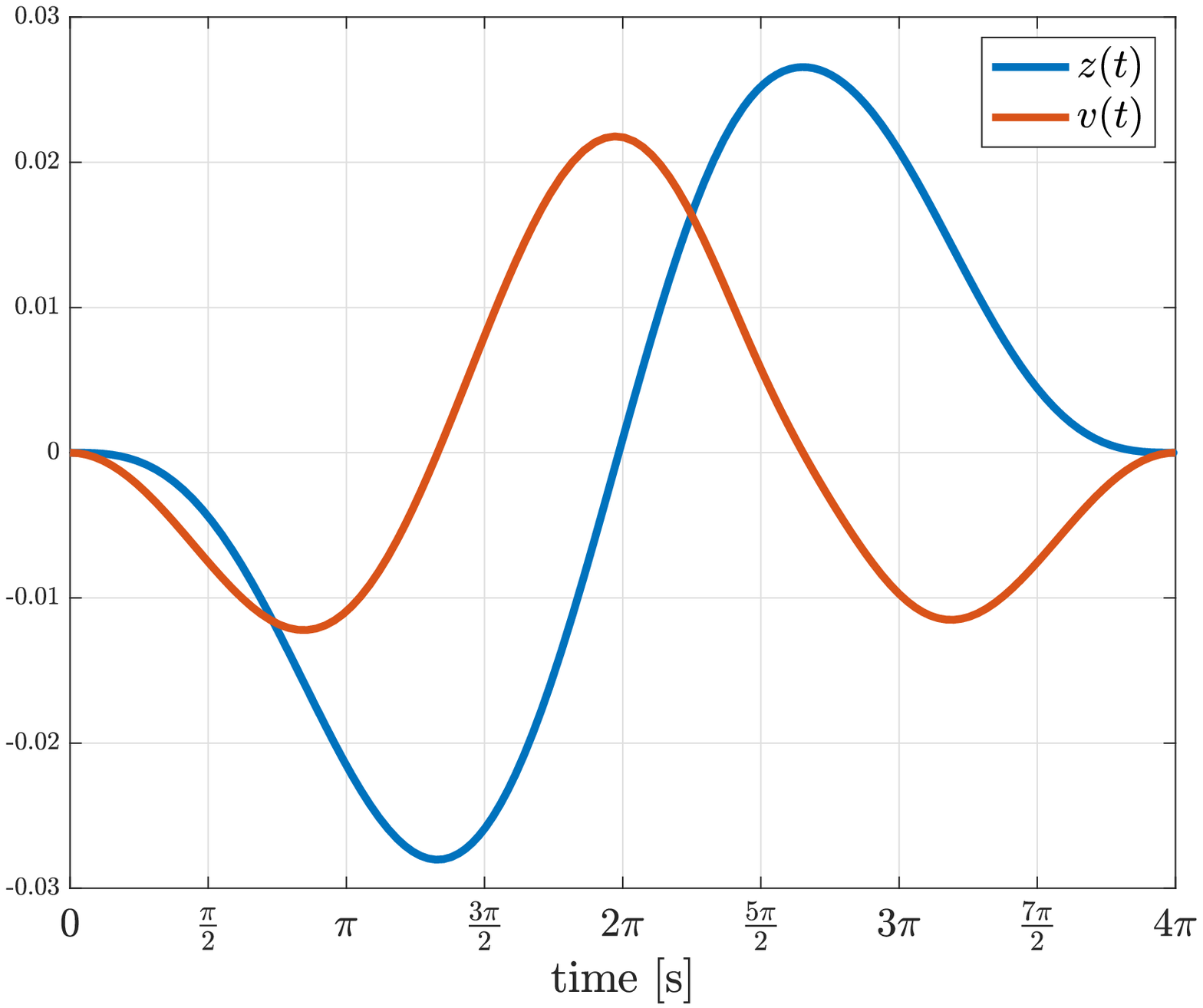}
\includegraphics[width=0.5\textwidth]{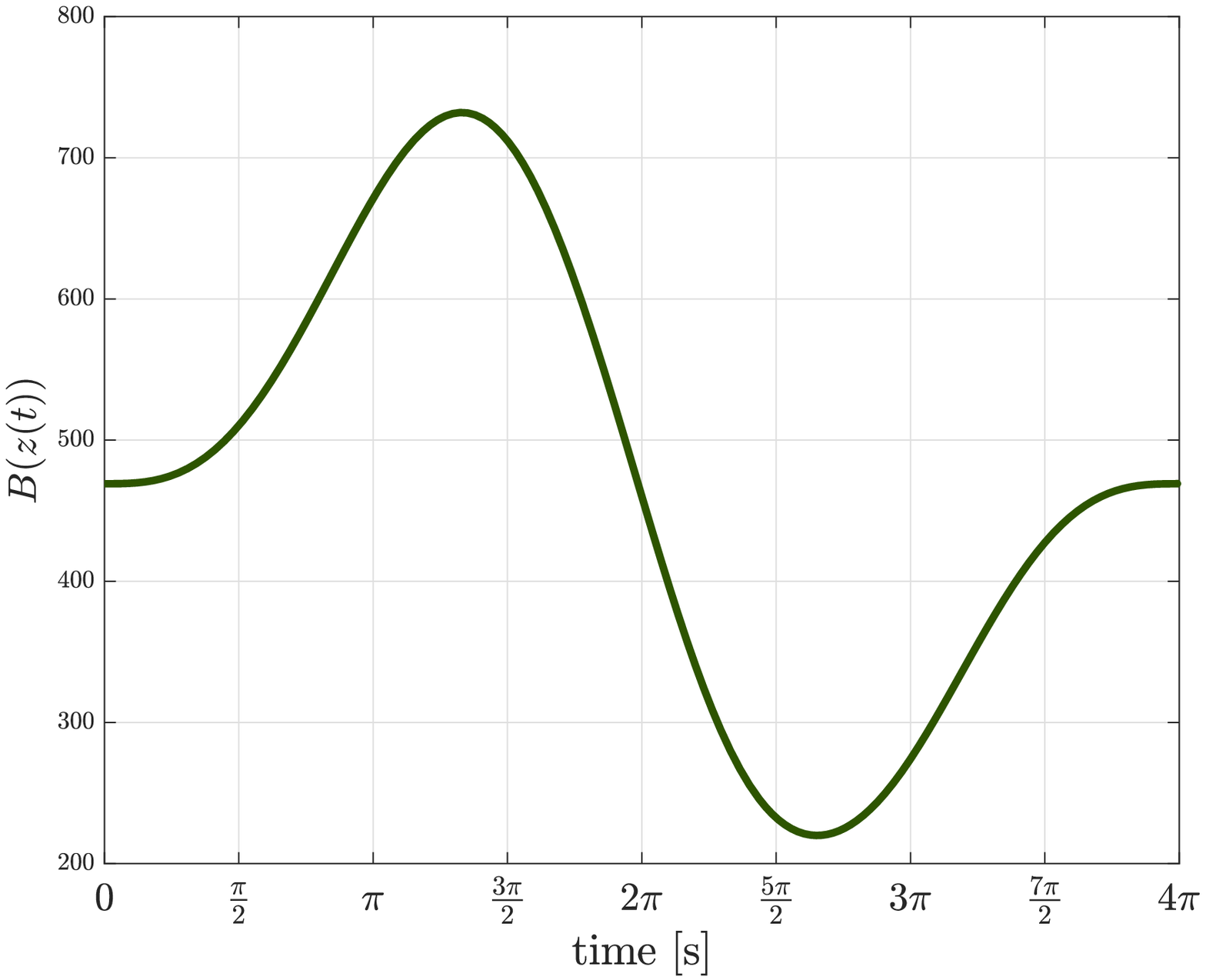}
\caption{Input, state, and ouput trajectories (top) and inertance (bottom).}
\label{fig:mem_inerter_iso}
\end{center}
\end{figure*}

\begin{figure*}[t]
\begin{center}
\includegraphics[width=0.8\textwidth]{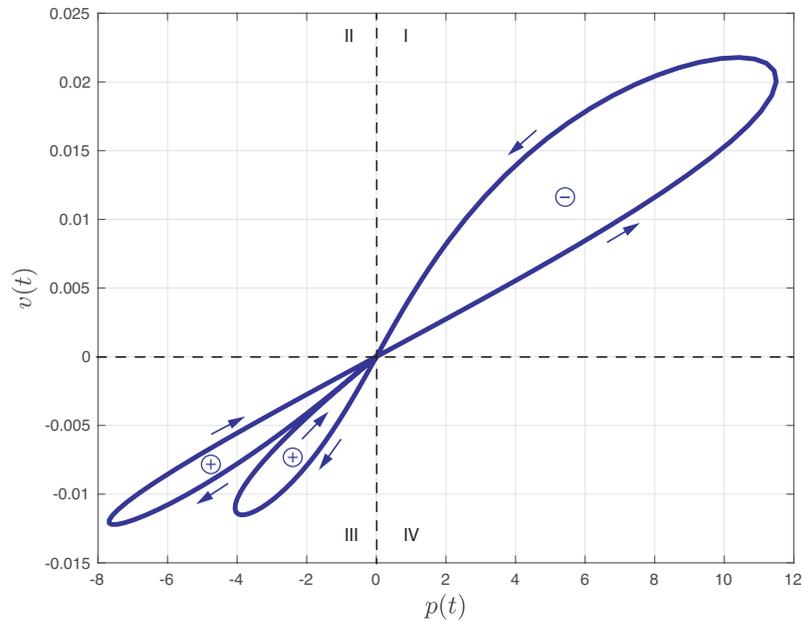}
\caption{Lissajous plot of the velocity versus the momentum.}
\label{fig:mem_inerter_lissa}
\end{center}
\end{figure*}

The simulation results are shown in Figure \ref{fig:mem_inerter_iso}. It is observed that both states start from and return to the same ground states, i.e., 
\begin{equation*}
z(0) = z(T) = 0 \ \text{[m]} \ \text{and} \ p(0) = p(T) = 0 \ \text{[Ns]}.
\end{equation*}
Interestingly, the supplied energy over one cycle equals
\begin{equation*}
\oint\limits_0^{T} v(t)F(t)dt = \boxed{-0.056 \ \text{[J]}}
\end{equation*}
which shows that there is more energy gained from the device than it is supplied with. Moreover, for each subsequent cycle $T$, the gained energy increases linearly with the number of cycles. This is also evident from the Lissajous plot in Figure \ref{fig:mem_inerter_lissa}; the loop in quadrant I has a counter-clockwise orientation (energy gain), whereas the loops in quadrant III both have a clockwise orientation (energy lost); the energy gained clearly exceeds the loss of energy. Thus, we can apparently harvest `free' energy from a mem-inerter, which confirms and exemplifies the assertions of Section \ref{sec:memcap_memind_passivity}. 

\section{Conclusion}\label{sec:conclusion}

It is shown in this paper that ideal memcapacitors, meminductors, and their possible analogies (like the recently proposed mem-inerter) constitute so-called  \emph{overunity} systems, Overunity systems belong to the class of perpetual motion machines of the first kind and are so far non-existing devices in the real-world as they are in direct conflict with the First Law of thermodynamics.\cite{kondepudi} For this reason alone, ideal real-world memcapacitors and meminductor will most probably never see the light of day. Of course, one could argue that for a certain class of inputs an ideal memcapacitor or meminductor might exhibit a (cyclo-)passive behavior. Purely sinusoidal inputs, which are most often used in the literature to reveal pinched hystereses loops, are an example of periodic excitations that generally yield cyclo-passive or even cyclo-lossless input-output behavior. However, for a device to be truly (cyclo-)passive, such behavior should be observable for \emph{all} possible excitation profiles! The use of ideal memcapacitors or meminductors as a realistic and reliable device model for mimicking real-world devices\cite{Chua2003,DiVentra2009,Radwan2015} is therefore highly debatable; 
\begin{center}
\it All (device) models are wrong, but some (device) models are more wrong than others. 
\end{center}

\end{document}